\newif{\ifjournal}
\newif{\ifarticle}
\journalfalse
\articlefalse

\newlength{\fsize}

\ifjournal
  \documentclass{aa}
  \usepackage{graphicx,times}
\else
  \ifarticle
    \documentclass[12pt]{article}
    \usepackage{a4wide,amssymb,graphicx,mathptm}
  \else
    \documentclass{paper}
  \fi
  \newcommand{\la}{\lesssim}
  \newcommand{\ga}{\gtrsim}
\fi

\renewcommand{\d}{\mathrm{d}}

\begin{document}

\ifjournal
  \title{Weak-Lensing Halo Numbers and Dark-Matter Profiles}
  \author{Matthias Bartelmann\inst{1} \and Lindsay J.~King\inst{1,2}
    \and Peter Schneider\inst{2}}
  \institute{Max-Planck-Institut f\"ur Astrophysik, P.O.~Box 1317,
    D--85741 Garching, Germany \and Institut f\"ur Astronomie und
    Extraterrestrische Forschung, Universit\"at Bonn, Auf dem H\"ugel
    71, D--53121 Bonn, Germany}
  \date{\today}

  \authorrunning{M.~Bartelmann, L.~King, P.~Schneider}
  \titlerunning{Weak-Lensing Halo Numbers}
\else
  \title{Weak-Lensing Halo Numbers and Dark-Matter Profiles}
  \author{Matthias Bartelmann$^{1}$, Lindsay J.~King$^{1,2}$, and
    Peter Schneider$^{2}$\\
    $^1$ Max-Planck-Institut f\"ur Astrophysik,
    \ifarticle\\\fi P.O.~Box 1317, D--85741 Garching, Germany\\
    $^2$ Institut f\"ur Astronomie und Extraterrestrische Forschung,
    Universit\"at Bonn,\\ Auf dem H\"ugel 71, 53121 Bonn}
  \date{\today}
  \ifarticle\maketitle\fi
\fi

\newcommand{\abstext}
  {Integral measures of weak gravitational lensing by dark-matter
   haloes, like the aperture mass, are sensitive to different physical
   halo properties dependent on the halo mass density profile. For
   isothermal profiles, the relation between aperture mass and virial
   mass is steeper than for haloes with the universal NFW
   profile. Consequently, the halo mass range probed by the aperture
   mass is much wider for NFW than for isothermal haloes. We use
   recent modifications to the Press-Schechter mass function in CDM
   models normalised to the local abundance of rich clusters, to
   predict the properties of the halo sample expected to be accessible
   with the aperture mass technique. While $\sim10$ haloes should be
   detected per square degree if the haloes have NFW profiles, their
   number density is lower by approximately an order of magnitude if
   they have isothermal profiles. These results depend only very
   mildly on the cosmological background model. We conclude that
   counts of haloes with a significant weak-lensing signal are a
   powerful discriminator between different dark-matter profiles.}

\ifjournal
  \abstract
    {\abstext
     }
\else
  \begin{abstract}
    \abstext
  \end{abstract}
\fi

\ifarticle\else\maketitle\fi

\section{Introduction\label{sec:1}}

One of the manifestations of gravitational lensing is the coherent
distortion of the images of faint background galaxies near foreground
matter concentrations due to their tidal gravitational field. This
effect is now well established for lensing by large-scale structure
[e.g. Van Waerbeke et al. 2000; Bacon, Refregier \& Ellis 2000;
Kaiser, Wilson \& Luppino 2000; Wittman et al.~2000; Maoli et
al.~2001; Van Waerbeke et al.~2001], in which case it is called
``cosmic shear'', and for lensing by galaxy clusters and
groups. Following pioneering work by Kaiser \& Squires (1993),
detailed two-dimensional mass maps of galaxy clusters were constructed
[e.g. Fischer \& Tyson 1997; Clowe et al.~2000; Hoekstra, Franx \&
Kuijken 2000] and the weaker effect of galaxy groups were stacked to
quantify their mean properties (Hoekstra et al.~2001). See for example
Mellier (1999) and Bartelmann \& Schneider (2001) for recent reviews
on weak lensing.

Being sensitive only to projected total mass, weak gravitational
lensing also allows the detection of sufficiently massive haloes only
by the effects of their mass, regardless of their composition,
physical state and relation to luminous material (Erben et al. 2000).
Cumulative measures for their gravitational tidal field, or shear,
were proposed, among which the so-called {\em aperture mass\/} (Kaiser
et al.~1994; Schneider 1996) is most directly related to
observations. The aperture mass is a suitably weighted integral of the
surface mass density inside a circular aperture, which is equivalent
to a differently weighted integral over the net tangential shear
within the aperture. The aperture mass is therefore a direct
observable.

It was recently shown in a different context (Bartelmann 2001) that it
depends on the density profile of a halo as to which {\em physical\/}
halo property the aperture mass actually measures. While the aperture
mass is $\propto M^{2/3}$ for singular isothermal profiles, it is
$\propto M^{1/3}$ for the universal profile proposed by Navarro, Frenk
\& White~(1997, hereafter NFW), with $M$ being the virial halo mass.

In this paper, we study the consequences of this difference in
detail. The basic idea is this. Given a fixed sensitivity limit of
aperture mass measurements, the flatter dependence of aperture mass on
virial mass for NFW haloes compared to isothermal haloes implies that
aperture mass measurements can cover a broader physical mass range if
the halo population has NFW rather than isothermal profiles. The
physical masses accessible to aperture mass measurements correspond to
rich groups and clusters, whose mass function is steep, turning over
from power-law to exponential decrease with increasing mass. Even a
small change in the accessible mass range can thus give rise to
substantial changes in the numbers of haloes expected to be detectable
with the aperture mass method. Conversely, this implies that number
counts of haloes with significant aperture mass can sensitively
constrain the halo density profiles, without the need to obtain
detailed density profiles of individual haloes.

The plan of the paper is as follows. We summarise the aperture mass
and its relevant properties in Sect.~\ref{sec:2}, and description of
the halo population in Sect.~\ref{sec:3}. Results are shown in
Sect.~\ref{sec:4}, and Sect.~\ref{sec:5} contains a discussion and our
conclusions.

\section{Aperture mass and halo profiles\label{sec:2}}

\subsection{Aperture mass}

The {\em aperture mass\/}, suggested in Schneider (1996) to quantify
the weak-lensing effects of dark-matter haloes, is an integral over
the lensing convergence $\kappa$ within a circular aperture of angular
radius $\theta$, weighted by a function $U(\vartheta)$ which vanishes
outside the aperture,
\begin{equation}
  M_\mathrm{ap}(\theta)=\int\d^2\vartheta\,\kappa(\vec\vartheta)\,
  U(|\vec\vartheta|)\;.
\label{eq:2.1}
\end{equation}
The prime advantage of $M_\mathrm{ap}$ is that it can directly be
determined from the measured tidal distortions of background-galaxy
images in the chosen aperture, provided $U(\vartheta)$ is compensated,
i.e.
\begin{equation}
  2\pi\,\int_0^\theta\,\d\vartheta\,\vartheta\,U(\vartheta)=0\;.
\label{eq:2.2}
\end{equation}
A broad class of weighting functions satisfies this condition. In
Schneider et al. (1998) the simple form was suggested,
\begin{equation}
  U(\vartheta)=\frac{9}{\pi\,\theta^2}\,\left[
    1-\left(\frac{\vartheta}{\theta}\right)^2
  \right]\,\left[
    \frac{1}{3}-\left(\frac{\vartheta}{\theta}\right)^2
  \right]
\label{eq:2.3}
\end{equation}
within the aperture, and $U(\vartheta)=0$ outside.

The aperture mass depends on source redshift because the convergence
$\kappa$ does. Since this is a linear dependence, and $M_\mathrm{ap}$
depends linearly on $\kappa$, a realistic source redshift distribution
like
\begin{equation}
  p(z_\mathrm{s})=\frac{\beta}{z_0^3\,\Gamma(3/\beta)}\,
  \exp\left[-\left(\frac{z}{z_0}\right)^\beta\right]
\label{eq:2.3a}
\end{equation}
can easily be taken into account by averaging $M_\mathrm{ap}$ over
$p(z_\mathrm{s})$. We choose $\beta=1.5$ and $z_0=1.0$ here and assume
from now on that source-redshift averaged aperture masses are being
used. For simplicity, we denote them by $M_\mathrm{ap}$ as before, as
there is no risk of confusion.

Schneider (1996) also calculated the dispersion of $M_\mathrm{ap}$ due
to the finite number of randomly distributed background galaxies and
their intrinsic ellipticities. Assuming typical values for the number
density $n_\mathrm{g}$ of suitably bright background galaxies and for
the dispersion $\sigma_\epsilon$ of their intrinsic ellipticities, it
was found that
\begin{eqnarray}
  \sigma_\mathrm{M}(\theta)&=&0.016\,
  \left(\frac{n_\mathrm{g}}{30\,\mathrm{arcmin}^2}\right)^{-1/2}\,
  \left(\frac{\sigma_\epsilon}{0.2}\right)\nonumber\\
  &\times&
  \left(\frac{\theta}{1\,\mathrm{arcmin}}\right)^{-1}\;.
\label{eq:2.4}
\end{eqnarray}

It can be shown by means of the Cauchy-Schwarz inequality that the
signal-to-noise ratio for a given halo is maximised if the weighting
function $U(\vartheta)$ is chosen such that it follows the halo
density profile. With the latter unknown, this cannot generally be
achieved, but we shall show later that the optimisation of weighting
functions is of minor importance.

The aperture size $\theta$ should be chosen large enough to encompass
a substantial number of background galaxies and to significantly cover
the extent of dark-matter haloes, and small enough in comparison with
typical field sizes and typical separations between neighbouring
haloes. Aperture sizes of a few arc minutes seem appropriate; to be
specific, we choose $\theta=3'$ in the following unless stated
otherwise [see also Kruse \& Schneider 1999].

\subsection{Mass profiles}

Let us now investigate the effect on the aperture mass if the density
profile of a halo of fixed physical mass $M$ is changed. We choose two
alternative density profiles, the singular isothermal sphere and the
profile suggested by NFW. We assume that halo masses are virial
masses, i.e.~masses confined by the virial radius $r_{200}$,
\begin{equation}
  r_{200}=\left(\frac{GM}{100\,H^2(z)}\right)^{1/3}\;,
\label{eq:2.5}
\end{equation}
where $H(z)$ is the Hubble function at redshift $z$. The virial radius
is defined such that the mean density within $r_{200}$ is $200$ times
the critical density.

\subsubsection{Singular isothermal spheres}

A singular isothermal sphere has the density profile
\begin{equation}
  \rho(r)=\frac{\sigma_v^2}{2\pi G\,r^2}\;,
\label{eq:2.6}
\end{equation}
where $\sigma_v$ is the (radially constant) velocity dispersion along
the line-of-sight. Integrating (\ref{eq:2.6}) out to the virial radius
(\ref{eq:2.5}), we find the virial mass
\begin{equation}
  M=\frac{2^{3/2}\,\sigma_v^3}{10G\,H(z)}\;.
\label{eq:2.7}
\end{equation}
It is important for the following that the velocity dispersion scales
with mass as
\begin{equation}
  \sigma_v\propto M^{1/3}\;.
\label{eq:2.8}
\end{equation}

Projecting the density profile (\ref{eq:2.6}) along the line-of-sight
and scaling the resulting surface mass density $\Sigma$ with its
critical value for lensing,
\begin{equation}
  \Sigma_\mathrm{cr}=\frac{c^2}{4\pi G}\,
  \frac{D_\mathrm{s}}{D_\mathrm{d}D_\mathrm{ds}}\;,
\label{eq:2.9}
\end{equation}
we arrive at the lensing convergence
\begin{equation}
  \kappa(\vartheta)=\frac{\theta_\mathrm{E}}{2\vartheta}\;,
\label{eq:2.10}
\end{equation}
where $\theta_\mathrm{E}$ is the angular Einstein radius,
\begin{equation}
  \theta_\mathrm{E}=4\pi\,\frac{\sigma_v^2}{c^2}\,
  \frac{D_\mathrm{ds}}{D_\mathrm{s}}\;.
\label{eq:2.11}
\end{equation}
Here and in (\ref{eq:2.9}), $D_\mathrm{d,s,ds}$ are the angular
diameter distances between observer and lens, observer and source, and
lens and source, respectively.

Inserting (\ref{eq:2.10}) and (\ref{eq:2.3}) into (\ref{eq:2.1}), we
find the aperture mass for the singular isothermal sphere,
\begin{equation}
  M_\mathrm{ap}^\mathrm{(SIS)}(\theta)=\frac{4}{5}\,
  \frac{\theta_\mathrm{E}}{\theta}\;.
\label{eq:2.12}
\end{equation}
Since the Einstein radius scales with $\sigma_v^2$, we see from
(\ref{eq:2.8}) that
\begin{equation}
  M_\mathrm{ap}^\mathrm{(SIS)}(\theta)\propto M^{2/3}\;.
\label{eq:2.13}
\end{equation}

\subsubsection{NFW haloes}

NFW used numerical $N$-body simulations to show that relaxed haloes
assume the universal density profile
\begin{equation}
  \rho(r)=\frac{\rho_\mathrm{crit}\,\delta_\mathrm{c}}
               {(r/r_\mathrm{s})\,(1+r/r_\mathrm{s})^2}\;,
\label{eq:2.14}
\end{equation}
where $\rho_\mathrm{crit}$ is the critical density and
$\delta_\mathrm{c}$ is a characteristic overdensity. The
characteristic radial scale $r_\mathrm{s}$ is related to the virial
radius through $r_\mathrm{s}=r_{200}/c$, where $c$ is the
concentration parameter. Bartelmann (1996) showed that the NFW profile
has the lensing convergence
\begin{equation}
  \kappa(x)=\frac{2\kappa_\mathrm{s}}{x^{2}-1}\,\left(
    1-\frac{2}{\sqrt{1-x^2}}\mathrm{arctanh}\sqrt{\frac{1-x}{1+x}}
  \right)\;,
\label{eq:2.15}
\end{equation}
with the convergence scale
\begin{equation}
  \kappa_\mathrm{s}\equiv
  \frac{\rho_\mathrm{crit}\delta_\mathrm{c}r_\mathrm{s}}
  {\Sigma_\mathrm{cr}}\;,
\label{eq:2.16}
\end{equation}
and $x\equiv r/r_\mathrm{s}$. NFW described how the parameters
$\delta_\mathrm{c}$ and $r_\mathrm{s}$ are related to the virial mass
$M$ of the halo. Hence, despite the two formal parameters in the
profile (\ref{eq:2.14}), it is entirely determined once the halo mass
is fixed. The statistics of dark-matter haloes described by the NFW
density profile has also been investigated by Kruse \& Schneider
(1999).

Unfortunately, there is no closed expression for the aperture mass of
an NFW profile. Numerically, however, it turns out to be fairly
shallow, approaching a constant for small apertures $\theta\to0$. It
is therefore appropriate to expand the aperture mass into a series in
\begin{equation}
  t\equiv\frac{\theta}{\theta_\mathrm{s}}\;,
\label{eq:2.17}
\end{equation}
were $\theta_\mathrm{s}$ is the angular scale radius,
$\theta_\mathrm{s}=r_\mathrm{s}/D_\mathrm{d}$. We find
\begin{eqnarray}
  M_\mathrm{ap}^\mathrm{(NFW)}(t)&\approx&\kappa_\mathrm{s}\,\left[
    1+t^2\left(\frac{19}{32}+\frac{3}{4}\ln\frac{t}{2}\right)
  \right.\nonumber\\
   &+&\left.
    t^4\left(\frac{77}{160}+\frac{3}{4}\ln\frac{t}{2}\right)
  \right]\;,
\label{eq:2.18}
\end{eqnarray}
which shows that $M_\mathrm{ap}^\mathrm{(NFW)}\to\kappa_\mathrm{s}$ in
the limit of small apertures. For haloes massive enough to produce a
significant weak-lensing signal,
$\rho_\mathrm{crit}\,\delta_\mathrm{c}$ varies only very little with
mass. Therefore, $M_\mathrm{ap}^\mathrm{(NFW)}$ essentially measures
the scale radius $r_\mathrm{s}$ or, equivalently,
\begin{equation}
  M_\mathrm{ap}^\mathrm{(NFW)}\propto M^{1/3}\;;
\label{eq:2.19}
\end{equation}
cf.~Eq.~(\ref{eq:2.5}).

Equations~(\ref{eq:2.19}) and (\ref{eq:2.13}) show that the {\em
aperture mass\/} measures different physical quantities depending on
the density profile of a given halo. Although the scaling with virial
mass is not wildly different between singular isothermal spheres and
NFW haloes, the effect on weak-lensing observations of a halo
population can be substantial, as we shall show below. For a given
virial halo mass, the virial radius is determined by
(\ref{eq:2.5}). Irrespective of the density profile, the halo mass
within $r_{200}$ is therefore fixed. For the isothermal profile, the
mass grows linearly with radius to reach $M$ at $r_{200}$, while it
grows more rapidly for $r\la r_\mathrm{s}$ and less rapidly beyond
$r_\mathrm{s}$ in case of an NFW density profile. The mass enclosed by
$r$ in an NFW halo thus exceeds that in a singular isothermal halo for
all $r\le r_{200}$. An NFW halo concentrates more mass in a given
radius $r\le r_{200}$, and therefore produces a stronger weak lensing
effect for a given aperture size if $r\ga r_\mathrm{s}$. Although the
previous scaling (\ref{eq:2.19}) was derived in the limit of small
apertures, $t\ll1$, the preceding argument shows that the main result
will continue to hold also for larger apertures, namely that an NFW
halo needs less mass to produce the same weak-lensing signal as a
singular isothermal halo.

\subsubsection{Weighting functions}

\begin{figure}[ht]
  \includegraphics[width=\fsize]{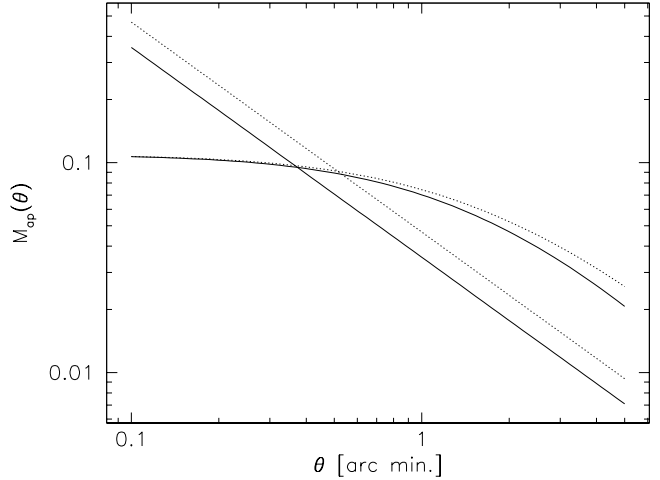}
\caption{Aperture mass profiles $M_\mathrm{ap}(\theta)$ for a singular
isothermal sphere (steeper curves) and an NFW halo (flatter curves)
with mass $10^{14}\,M_\odot/h$ at redshift $0.2$. In this figure, the
source redshift is $1.0$. While $M_\mathrm{ap}$ falls like
$\theta^{-1}$ for the SIS, it flattens for small $\theta$ for the NFW
case. The solid curves were obtained with the weighting function
(\ref{eq:2.3}), while (\ref{eq:2.19a}) was used to obtain the dotted
curves. The optimised weighting function (\ref{eq:2.19a}) improves
$M_\mathrm{ap}$ by $\sim30\%$ for the singular isothermal sphere, and
less for the NFW profile.}
\label{fig:1}
\end{figure}

As mentioned before, $M_\mathrm{ap}$ can be optimised by choosing the
weight function $U(\vartheta)$ such that it follows the halo density
profile (Schneider 1996). Of course, this cannot completely be
achieved because of the requirement (\ref{eq:2.2}) that $U(\vartheta)$
be compensated. It is, however, possible to choose the weighting
function such that it approximately follows the slope of the halo
density profile within most of the aperture. For singular isothermal
spheres, such an approximation could be
\begin{equation}
  U(\vartheta)\propto
  \frac{1}{\sqrt{\vartheta^2+\vartheta_\mathrm{c}^2}}-C\;,
\label{eq:2.19a}
\end{equation}
where $\vartheta_\mathrm{c}$ removes the central singularity, $C$ must
be chosen to satisfy (\ref{eq:2.2}) and the normalisation is the same
as for (\ref{eq:2.3}). For the NFW profile, an optimised weighting
function is more difficult to construct because the physical scale
$r_\mathrm{s}$ changes its angular size with halo distance, so that
the weighting function would have to vary with distance. Near
$r_\mathrm{s}$, where the NFW profile gently steepens from $r^{-1}$ to
$r^{-3}$, the profile is approximately isothermal, so some improvement
from using (\ref{eq:2.19a}) instead of (\ref{eq:2.3}) is also expected
for the NFW profile. Figure~\ref{fig:1} shows the aperture mass as a
function of aperture size for singular isothermal spheres and NFW
haloes, using the weighting functions (\ref{eq:2.3}) and
(\ref{eq:2.19a}).

The figure summarises the results obtained in this subsection. While
$M_\mathrm{ap}(\theta)$ falls $\propto\theta^{-1}$ for the singular
isothermal sphere, it flattens for $\theta\to0$ for the NFW
profile. The weighting function (\ref{eq:2.19a}), which is adapted to
the isothermal profile, improves $M_\mathrm{ap}(\theta)$ quite
appreciably for the singular isothermal sphere, and less so for the
NFW profile. Since the noise (\ref{eq:2.4}) is proportional to
$\theta^{-1}$, the S/N of $M_\mathrm{ap}(\theta)$ is constant with
$\theta$ for the singular isothermal sphere, and is maximised for the
NFW profile at a filter scale of a few arc minutes, where its slope is
isothermal.

\subsection{Halo substructure and asymmetry}

Since massive haloes are cosmologically young objects, they frequently
show substructure or deviations from axial symmetry. One may wonder
whether the relations derived above between aperture mass and virial
mass cease to hold if the halo symmetry is perturbed.

\begin{figure}[ht]
  \includegraphics[width=\fsize]{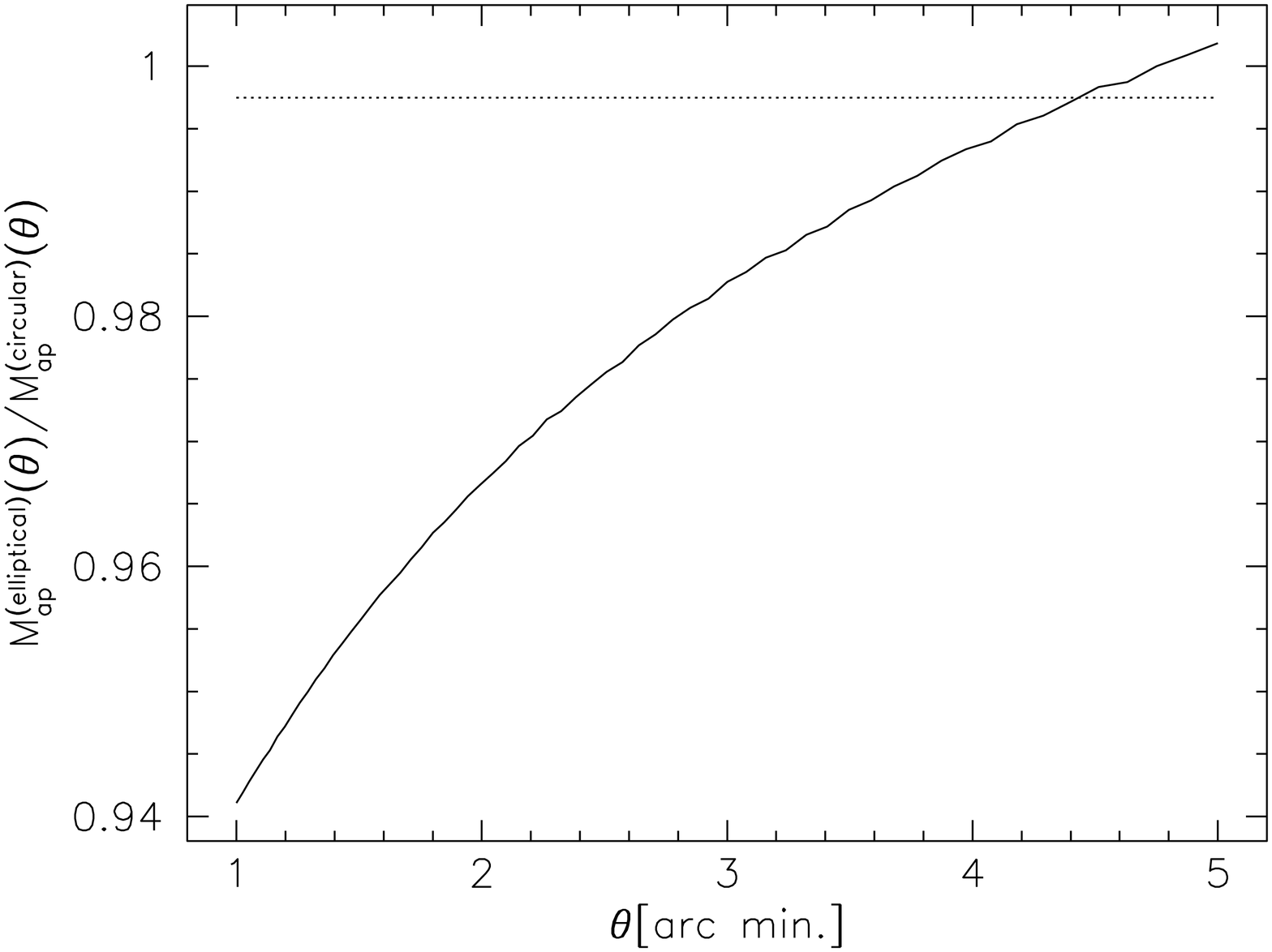}
\caption{Influence of halo ellipticity on the aperture mass. The solid
curve shows the ratio of the aperture masses for an elliptically
distorted NFW halo and one that is circularly symmetric. The assumed
ellipticity is $e=0.2$. At an aperture size of $\theta=3'$, the ratio
deviates from unity by less than $2$ per cent. The dotted curve shows
the same ratio for singular isothermal profiles, which is independent
of aperture size. Here, the deviation from unity is $0.25$ per cent.}
\label{fig:0}
\end{figure}

For a specific example, let us distort the projected NFW profile
(\ref{eq:2.15}) such that lines of constant surface density become
elliptical, while keeping the azimuthally-averaged density profile
unchanged. This is achieved by multiplying $\kappa(x)$ with
$(1-e)^{1/2}$ and replacing the radial distance $x$ by
\begin{equation}
  \xi=[x_1^2+(1-e)^2\,x_2^2]^{1/2}\;,
\label{eq:2.20}
\end{equation}
where $0\le e<1$ is the ellipticity parameter. To first order in $e$,
this changes the NFW convergence to
\begin{equation}
  \tilde{\kappa}(\xi)=\kappa(x)-e\,\left\{
    \frac{\kappa(x)}{2}+\frac{x_2^2}{1-x^2}\,
    \left[3\kappa(x)-\frac{2}{x^2}\right]
  \right\}\;.
\label{eq:2.22}
\end{equation}
The solid line in Fig.~\ref{fig:0} shows the ratio between the
aperture masses of the elliptically distorted and the circularly
symmetric NFW profiles, $\tilde{\kappa}(\xi)$ and $\kappa(x)$, for
$e=0.2$.

The influence of halo ellipticity can also be quantified by comparing
$M_\mathrm{ap}^\mathrm{(SIE)}$ for the singular isothermal ellipsoid
(Kormann et al.~1994) with $M_\mathrm{ap}^\mathrm{(SIS)}$. The
convergence of the SIE is
\begin{equation}
  \kappa(\vartheta,\phi)=\frac{\theta_\mathrm{E}}{2\vartheta}\,
  \frac{\sqrt{1-e}}{\sqrt{\cos^{2}\phi+(1-e)^{2}\sin^{2}\phi}}\;,
\label{eq:L1}
\end{equation}
where $\phi$ is the position angle with respect to its major axis and
$\theta_\mathrm{E}$ is the equivalent circular Einstein radius. Note
that this reduces to the SIS when $e=0$. We find the closed form
\begin{equation}
  \frac{M_\mathrm{ap}^\mathrm{(SIE)}}{M_\mathrm{ap}^\mathrm{(SIS)}}=
  \frac{2\sqrt{1-e}}{\pi}K\left(\sqrt{1-(1-e)^{2}}\right)\;,
\label{eq:L2}
\end{equation}
where $K(m)$ is the complete elliptic integral of the first kind,
following the notation in Gradshteyn \& Ryzhik (1965). There is no
dependence upon the particular form of filter function adopted, or
upon its scale. The ratio (\ref{eq:L2}) deviates from unity by less
than 3 per cent if $e\le0.5$, and by less than 11 per cent if
$e\le0.75$. A Taylor expansion for (\ref{eq:L2}), to second order in
$e$ about $e=0$ reads
\begin{equation}
  \frac{M_\mathrm{ap}^\mathrm{(SIE)}}{M_\mathrm{ap}^\mathrm{(SIS)}}=
  1-\frac{e^{2}}{16}+\mathcal{O}(e)^{3}\;.
\label{eq:L3}
\end{equation}
For $e=0.2$, this deviates from unity by $0.25$ per cent, indicated by
the dotted line in Fig.~\ref{fig:0}.


Now consider how substructure changes the measured aperture mass, and
how the aperture mass obtained for a ``realistic'' cluster deviates
from $M_\mathrm{ap}^\mathrm{(NFW)}$. Intuitively, the presence of
substructure in a halo should not influence the aperture mass derived
from $\kappa$ provided that the total halo mass remains fixed. The
most accurate description of substructure is obtained from high
resolution $N$-body simulations, and we use a scaled cluster from
Springel (1999); a preliminary consideration of $M_\mathrm{ap}$
applied to such simulations was presented in King et al.~(2000). For
consistency with the simulations we adopt $h=0.7$, $\Omega_{0}=0.3$
and $\Omega_{\Lambda}=0.7$. The value of
$M_\mathrm{ap}^\mathrm{(SIM)}$ obtained for this simulated cluster is
the same as that of its azimuthally averaged counterpart, within
numerical accuracy.

\begin{figure}[ht]
  \includegraphics[width=\fsize]{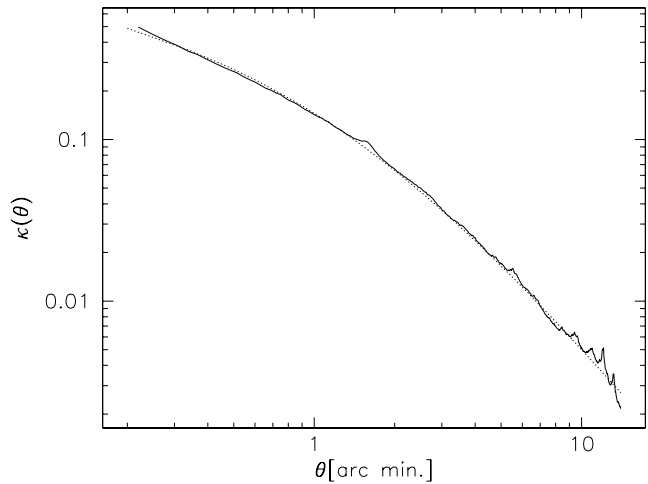}
\caption{The true convergence profile of the simulated cluster with
substructure (solid line) and the best-fit NFW convergence (dotted
line).}
\label{fl2}
\end{figure}

Figure~\ref{fl2} shows the convergence of the simulated cluster, and
of the best-fit NFW model ($c=4.9$, $r_{200}=1.1~{\rm Mpc}$) over the
range plotted. Towards smaller radii, the simulation profile is
steeper than NFW, resulting in the best-fit NFW profile having a
larger concentration parameter when the inner fitting radius is
decreased.  The differences between the aperture masses
$M_\mathrm{ap}^\mathrm{(SIM)}$ calculated directly from the cluster
simulation and $M_\mathrm{ap}^\mathrm{(NFW)}$ for the best-fit NFW
model are also negligible for aperture sizes of interest. In other
words, the presence of substructure at a level consistent with
$N$-body simulations does not significantly change the relationships
between aperture mass and virial mass.

\section{Halo population\label{sec:3}}

An approximate description for the mass function of haloes was given
by Press \& Schechter (1976). In terms of mass $M$ and redshift $z$,
their mass function can be written as
\begin{eqnarray}
  n_\mathrm{PS}(M,z)&=&\frac{\bar\rho}{\sqrt{2\pi}\,D_+(z)\,M^2}\,
  \left(1+\frac{n}{3}\right)\,
  \left(\frac{M}{M_\ast}\right)^{(n+3)/6}\nonumber\\
  &\times&\exp\left[-\frac{1}{2\,D_+^2(z)}\,
  \left(\frac{M}{M_\ast}\right)^{(n+3)/3}\right]\;,
\label{eq:3.1}
\end{eqnarray}
where $M_\ast$ and $\bar\rho$ are the nonlinear mass today and the
mean background density {\em at the present epoch\/}, and $D_+(z)$ is
the linear growth factor of density perturbations, normalised to unity
today, $D_+(0)=1$. Finally, $n$ is the effective exponent of the
dark-matter power spectrum at the cluster scale, $n\approx-1$.

Sheth \& Tormen (1999) recently modified the mass function
(\ref{eq:3.1}), and Sheth et al.~(1999) introduced ellipsoidal rather
than spherical collapse. Jenkins et al.~(1999) derived the mass
function of dark-matter haloes from numerical simulations and found a
fitting formula very close to Sheth \& Tormen's, but with lower
amplitude at the high-mass end. Although these different mass
functions are very similar over a wide range of masses, their
different behaviour at high mass can lead to noticeable changes in our
results. We use the mass function by Jenkins et al.~for the results
shown below.

For rich groups and clusters, the halo mass function is steep in
mass. Therefore, even a moderate change in the mass range considered
can lead to substantial changes in the total halo number. We saw above
that the aperture mass effectively measures different fractional
powers of the halo mass depending on the density profile. As the halo
mass decreases, $M_\mathrm{ap}$ decreases faster for singular
isothermal spheres than for haloes with NFW profile. The halo mass
range probed by the aperture mass will therefore be narrower if the
halo population consists of singular isothermal spheres than if it
consists of NFW haloes. Considering the steepness of the halo mass
function, we expect to see substantially more haloes with significant
aperture mass if the halo population is characterised by the NFW
density profile compared to the singular isothermal profile. We will
now quantify this expectation.

\section{Results\label{sec:4}}

We describe significant weak gravitational lensing by a halo in terms
of the signal-to-noise ratio,
\begin{equation}
  \mathcal{S}(\theta)=
  \frac{M_\mathrm{ap}(\theta)}{\sigma_\mathrm{M}(\theta)}\;,
\label{eq:4.1}
\end{equation}
with $\sigma_\mathrm{M}(\theta)$ given by (\ref{eq:2.4}). To be
specific, we require $\mathcal{S}\ge\mathcal{S}_\mathrm{min}$ for a
significant measurement of the aperture mass, and choose
$\mathcal{S}_\mathrm{min}=5$ in the following.

In the absence of noise, there would be a one-to-one relation between
aperture mass and physical mass, so that a sharp detection cut-off on
$M_\mathrm{ap}$ would map onto a sharp threshold in mass. However,
since the aperture-mass dispersion (\ref{eq:2.4}) is finite, some
haloes with mass below the threshold have higher, and some haloes with
mass above the threshold have lower aperture masses than the
cut-off. With the halo mass function falling steeply, on average more
low-mass haloes are gained in the sample than high-mass haloes
lost. We must therefore take this bias into account, which we do by
convolving the sharp boundary in mass with a Gaussian of width
\begin{equation}
  \Delta M=\left(
    \frac{\partial\mathcal{S}}{\partial M}
  \right)^{-1}\;.
\label{eq:4.1a}
\end{equation}
Needless to say, $\Delta M$ depends on halo mass and redshift.

\begin{figure}[ht]
  \includegraphics[width=\fsize]{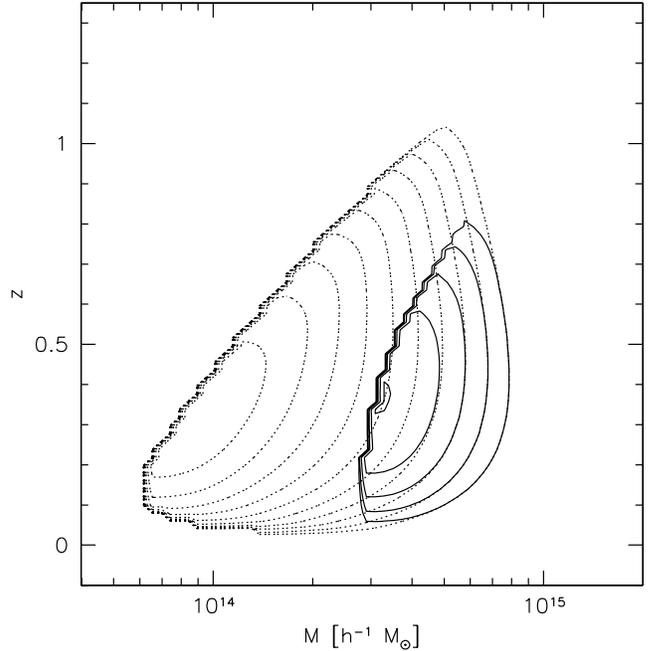}
\caption{Number density of haloes $\d^2N/\d M\d z$ in the $M$-$z$
plane, integrated over the sky, which produce a significant
weak-lensing signal. Solid contours: singular isothermal spheres,
dashed contours: haloes with NFW density profile. Evidently, the
distribution of gravitationally lensing haloes in mass and redshift
changes substantially with the density profile. The contour levels
range from $10^{-10.5}$ to $10^{-7}\,h\,M_\odot^{-1}$ and are spaced
by $0.25$ dex. The cosmology is $\Lambda$CDM ($\Omega_{0}=0.3$,
$\Omega_{\Lambda}=0.7, h=0.7$).}
\label{fig:2}
\end{figure}

Figure~\ref{fig:2} shows contours of the number density in the
mass-redshift plane of haloes with a significant aperture mass,
integrated over the sky. The underlying cosmological model is
spatially flat and has low density, $\Omega_0=0.3$,
$\Omega_\Lambda=0.7$, $h=0.7$, and the dark-matter power spectrum is
normalised such that the local abundance of rich clusters is
reproduced. The solid contours are for singular isothermal spheres,
the dotted contours for NFW haloes. Obviously, the mass range in which
NFW haloes can produce a significant weak-lensing signal is much wider
than for singular isothermal spheres.

\begin{figure*}[ht]
  \includegraphics[width=\fsize]{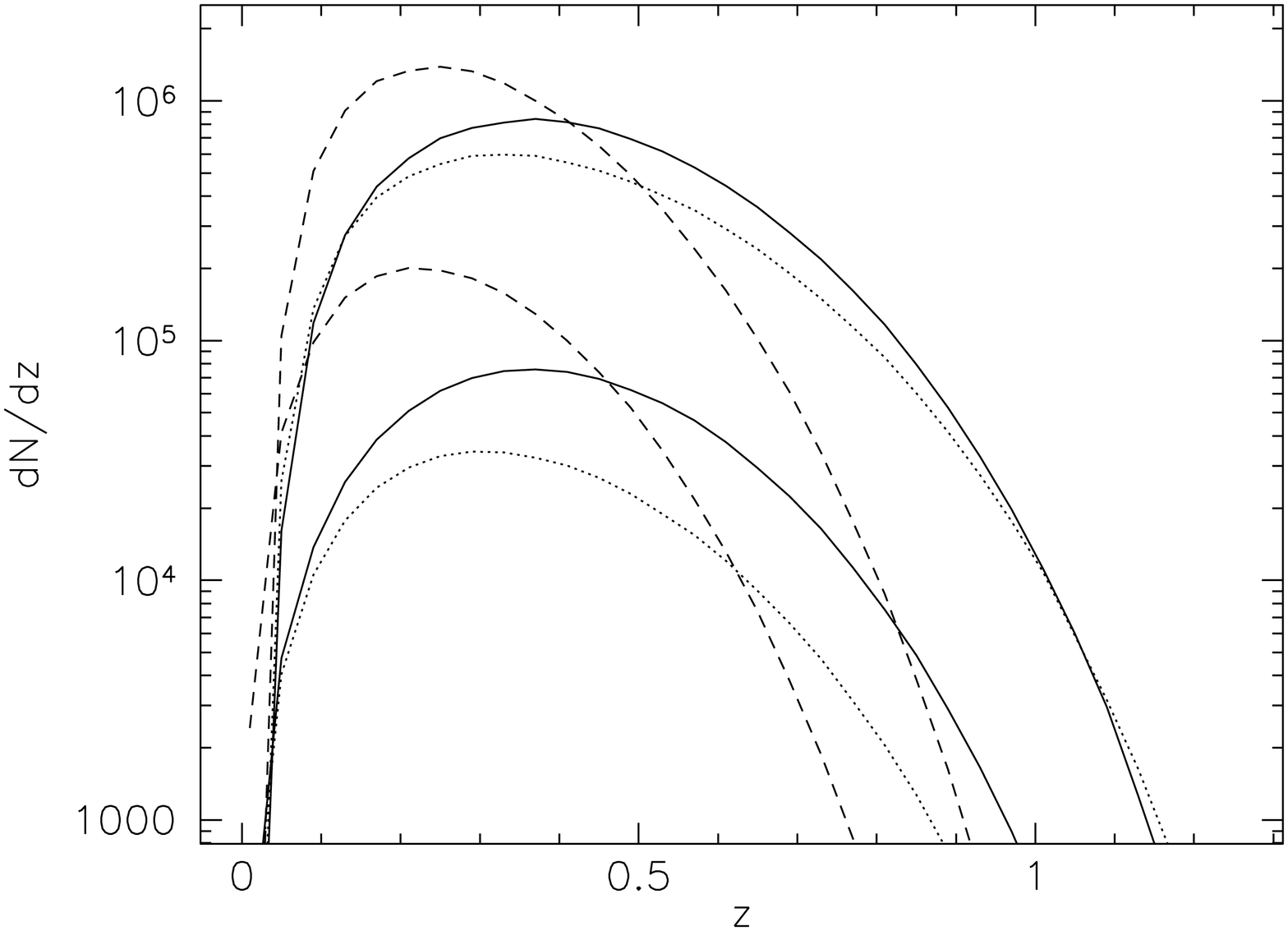}\hfill
  \includegraphics[width=\fsize]{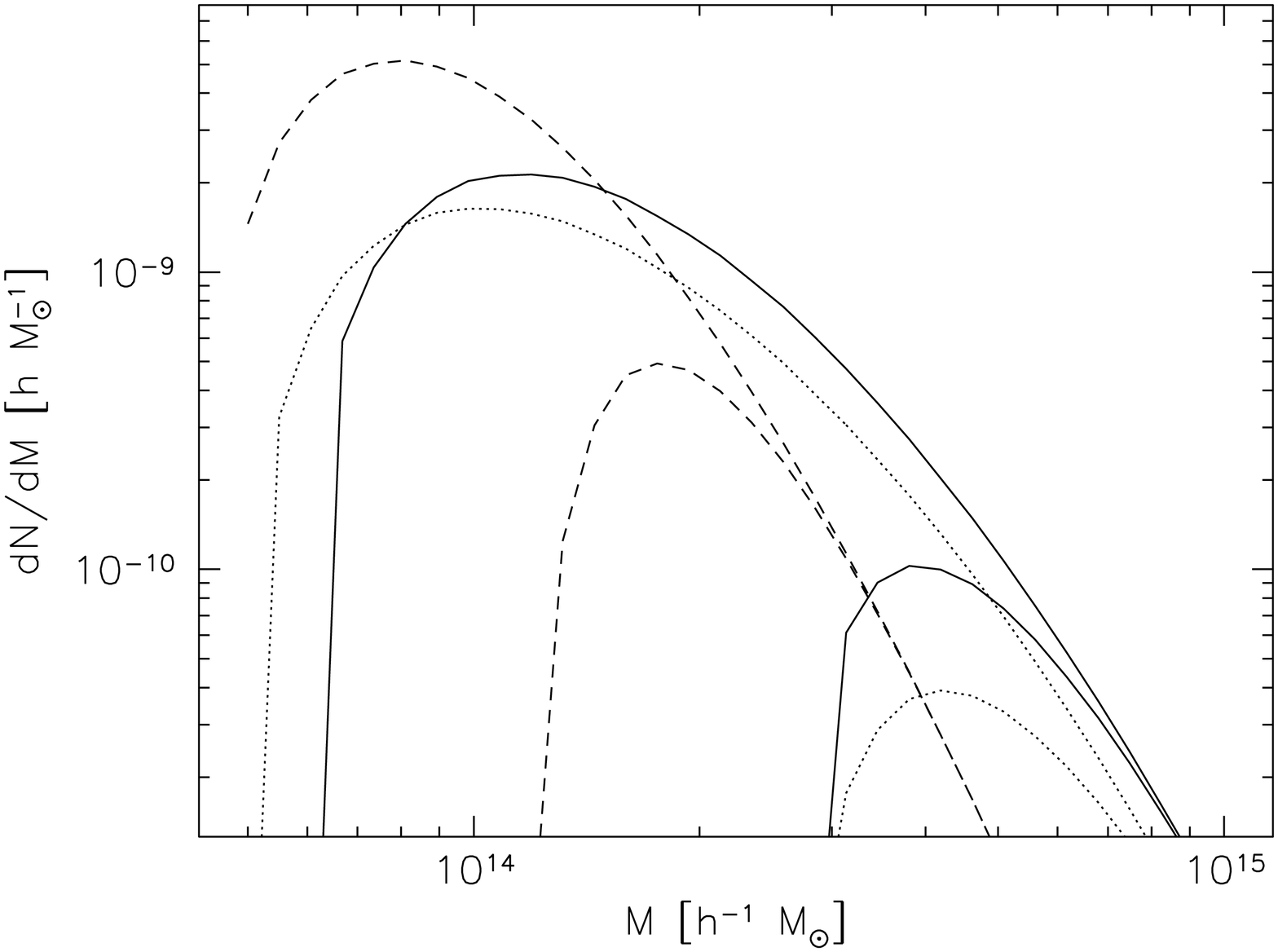}
\caption{Distribution of gravitationally lensing haloes in redshift
(left panel) and mass (right panel) for three different cosmological
models and two different density profiles. Solid curves: $\Lambda$CDM,
dotted curve: OCDM, dashed curve: SCDM. In both panels, the three
curves with the lower amplitude refer to SIS haloes, those with the
higher amplitude to NFW haloes. While the redshift range is fairly
insensitive to the density profile, the mass range reaches to
substantially lower mass for a halo population with NFW density
profile compared to a population with singular isothermal profile.}
\label{fig:3}
\end{figure*}

While Fig.~\ref{fig:2} qualitatively illustrates the effect expected,
the two panels of Fig.~\ref{fig:3} show the distribution of
weak-lensing haloes in redshift and mass, respectively. Three pairs of
curves are shown in each figure for three representative cosmological
models, all of which are normalised to reproduce the local cluster
abundance. The models are $\Lambda$CDM ($\Omega_0=0.3$,
$\Omega_\Lambda=0.7$, $h=0.7$), OCDM ($\Omega_0=0.3$,
$\Omega_\Lambda=0.0$, $h=0.7$), and SCDM ($\Omega_0=1.0$,
$\Omega_\Lambda=0.0$, $h=0.5$). Quite independent of the cosmological
model, the peak amplitudes of the curves for singular isothermal
spheres are approximately one order of magnitude lower than for NFW
haloes.

\begin{figure}[ht]
  \includegraphics[width=\fsize]{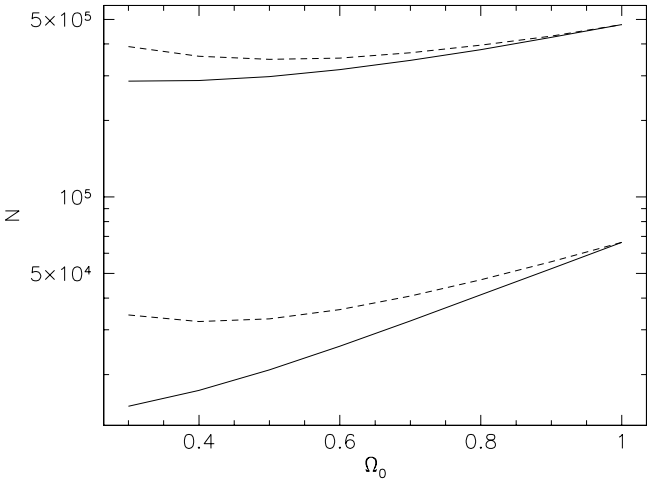}
\caption{Total number of gravitationally lensing haloes on the sky as
a function of the cosmic density parameter $\Omega_0$. Solid curves:
$\Omega_\Lambda=0$, dashed curves: $\Omega_\Lambda=1-\Omega_0$. The
lower pair of curves refers to SIS haloes, the upper to NFW haloes.}
\label{fig:5}
\end{figure}

Figure~\ref{fig:5} shows the expected number on the sky of haloes with
a significant aperture mass, in dependence on the density parameter
$\Omega_0$. Two pairs of curves are shown. The upper pair refers to
NFW haloes, the lower one to singular isothermal spheres. In each
pair, the solid curve was calculated for open universes,
$\Omega_\Lambda=0$, and the dashed curve for spatially flat universes,
$\Omega_\Lambda=1-\Omega_0$. Note that the Hubble constant is not
varied here, $h=0.7$.

On the whole, the expected number of weak-lensing haloes is higher by
about an order of magnitude than if the haloes have an NFW rather than
a singular isothermal density profile. The dependence on cosmology is
very mild, in particular for the NFW haloes, and its trend seems
counter-intuitive. Haloes tend to form earlier in low- than in
high-density universes, hence one should expect the number of
weak-lensing haloes to increase with decreasing $\Omega_0$. Contrary
to that expectation, the halo number tends to increase slightly with
increasing $\Omega_0$.

\begin{figure}[ht]
  \includegraphics[width=\fsize]{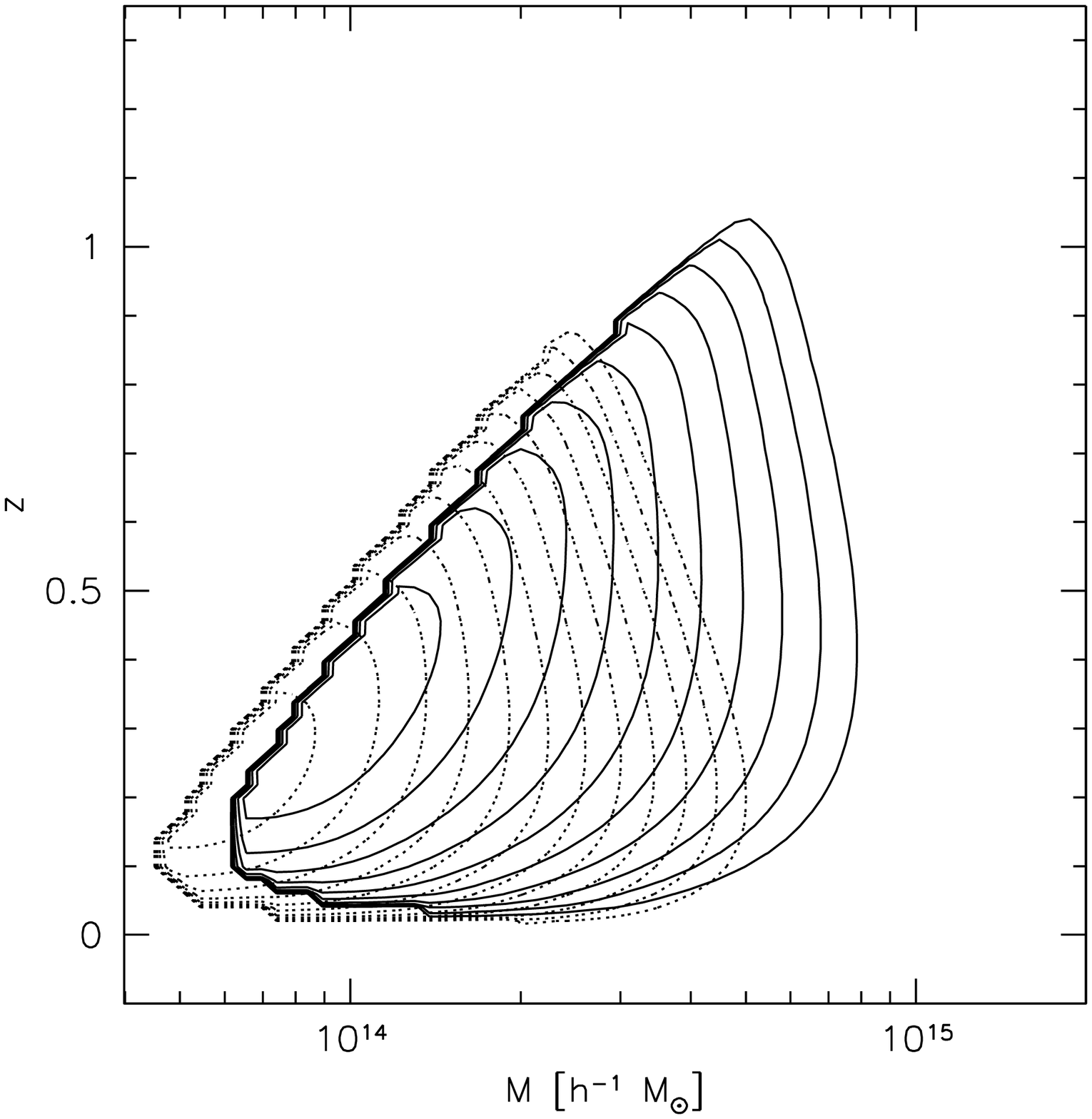}
\caption{Similar to Fig.~\ref{fig:2}, contours are shown of the number
over all the sky of haloes in the $M$-$z$ plane which produce a
significant aperture mass. The haloes are assumed to have NFW density
profiles. The solid and dotted sets of contours refer to the
$\Lambda$CDM and SCDM models, respectively. As in Fig.~\ref{fig:2},
the contour levels range from $10^{-10.5}$ to
$10^{-7}\,h\,M_\odot^{-1}$ and are spaced by $0.25$ dex. For high
$\Omega_0$, the contours shift to lower mass because haloes in
high-density universes tend to be more compact, as described in the
text.}
\label{fig:7}
\end{figure}

The reason for this trend is the fact that the overdensity inside
virialised haloes grows with increasing $\Omega_0$. This is most
easily seen in the spherical collapse model, in which the virial
overdensity for critical-density universes is
$\Delta_\mathrm{c}=18\pi^2\approx178$ independent of redshift, but
lower by a factor of $\sim2$ for $\Omega_0=0.3$ at low
redshifts. Consequently, haloes of fixed mass have higher
$M_\mathrm{ap}$ in high-density universes, or, conversely, haloes that
produce a given minimum $M_\mathrm{ap}$ can have lower mass in
high-density universes. We have seen before that small changes in the
mass range probed lead to large changes in the halo numbers because
the mass function is so steep. Hence, the higher compactness of haloes
in high-density universes leads to an increase in the number of {\em
visible\/} haloes that overcompensates for the more rapid decrease in
halo number with increasing redshift. Figure~\ref{fig:7} further
illustrates this point.

\section{Discussion and Conclusions\label{sec:5}}

We studied in this paper how the number of dark-matter haloes with a
significant weak-lensing signal depends on the halo density
profile. We modelled the halo population following the modification to
the Press-Schechter mass function by Jenkins et al.~(2000), and their
density profiles as either singular isothermal or NFW profiles. As a
measure for weak gravitational lensing, we chose the aperture mass,
which is a weighted integral over the scaled surface mass density
within a circular aperture. Due to its direct relation to the
gravitational tidal field, the aperture mass $M_\mathrm{ap}$ is a
directly observable quantity. We call the expected weak lensing signal
significant if its signal-to-noise ratio is five or higher.

We further assume that the dark matter has a CDM power spectrum,
normalised such that the local abundance of rich galaxy clusters is
reproduced.

Our results can be summarised as follows:

\begin{enumerate}

\item For singular isothermal spheres, aperture mass $M_\mathrm{ap}$
and physical mass $M$ are related by $M_\mathrm{ap}\propto M^{2/3}$,
for NFW haloes, by $M_\mathrm{ap}\propto M^{1/3}$. This result was
found earlier (Bartelmann 2001), but we repeat it here for
completeness. Depending on the density profile, the aperture mass is
therefore sensitive to different scalings of the physical halo mass.

\item Changes in the relation between $M_\mathrm{ap}$ and $M$ imply
that the physical halo mass range probed by $M_\mathrm{ap}$ depends on
the density profile. It is wider if the relation is flatter, hence
$M_\mathrm{ap}$ probes a larger physical mass range if the haloes have
NFW profiles.

\item Weak-lensing measures like $M_\mathrm{ap}$ are sensitive to
haloes with masses $M\ga5\times10^{13}\,h^{-1}\,M_\odot$, above the
non-linear mass today,
$M_\ast\sim10^{13}\,h^{-1}\,M_\odot$. Therefore, in the mass range
probed by weak lensing, the halo mass function steepens from the
power-law to the exponential fall-off. Even moderate extensions of the
mass range therefore lead to substantial changes in the detectable
halo numbers.

\item The mass range probed in low-density universes
($\Omega_0\sim0.3$) is $\sim(1.5\pm0.9)\times10^{14}\,h^{-1}\,M_\odot$
for NFW haloes, and notably shifted to higher masses for singular
isothermal haloes, $\sim(4.8\pm1.5)\times10^{14}\,h^{-1}\,M_\odot$. In
high-density universes ($\Omega_0\sim1.0$), typical masses are lower;
$\sim(1.0\pm0.5)\times10^{14}\,h^{-1}\,M_\odot$ for NFW, and
$\sim(2.1\pm0.7)\times10^{14}\,h^{-1}\,M_\odot$ for singular
isothermal haloes.

\item For low-density universes, the redshift range of the
weak-lensing haloes is $\sim(0.4\pm0.18)$, while it is slightly
narrower and shifted to somewhat lower redshift, $\sim(0.3\pm0.14)$,
for high-density universes. The redshift range is fairly independent
of the halo density profile.

\item Quite independent of the cosmological parameters, the expected
number of significantly lensing haloes per square degree is $\sim10$
if the haloes have NFW density profiles, and approximately {\em an
order of magnitude less\/} if they are singular isothermal
spheres. The somewhat surprising result that the number of
weak-lensing haloes does not substantially change with cosmic density
is due to the balance between halo evolution and halo
compactness. While the number of haloes decreases much more rapidly
with increasing redshift in a high-density universe, their virial
overdensity is higher, giving rise to a stronger lensing signal. As
illustrated in Fig.~\ref{fig:7}, this allows the detection of haloes
with somewhat smaller mass, which significantly increases the
detectable halo number because of the steepness of the mass function.

\end{enumerate}

These results demonstrate the possibility to constrain the density
profile of dark matter haloes by counting how many haloes per square
degree produce a significant weak lensing signal. Of course,
dark-matter density profiles can also be directly measured using weak
lensing methods. However, such measurements suffer from the intrinsic
resolution limit of weak lensing due to the finite number of
background galaxies, and the fact that the NFW density profile has
approximately isothermal slope in the radial range where weak-lensing
measures are most sensitive. Furthermore, King \& Schneider~(2001)
investigated how easy it is to distinguish between NFW and power-law
profiles, using a maximum likelihood approach. It was concluded that
wide field images are required to put a significant constraint on the
density profile of an individual cluster.

\ifjournal\begin{acknowledgements}\else\section*{Acknowledgements}\fi
This work was supported by the TMR Network ``Gravitational Lensing:
New Constraints on Cosmology and the Distribution of Dark Matter'' of
the EC under contract No. ERBFMRX-CT97-0172.  We would like to thank
Volker Springel and Simon White for very kindly allowing us to use
their cluster simulations, and Houjun Mo for his careful reading of
the manuscript.
\ifjournal\end{acknowledgements}\fi

\end{document}